# Coarse Molecular-Dynamics Determination of the Onset of Structural Transitions: Melting of Crystalline Solids

Miguel A. Amat[1], Ioannis G. Kevrekidis[2], and Dimitrios Maroudas[1,*]

[1] Department of Chemical Engineering, University of Massachusetts, Amherst, MA 01003
[2] Department of Chemical Engineering and Program in Applied and Computational Mathematics, Princeton University, Princeton, NJ 08544

Using a coarse molecular-dynamics (CMD) approach with an appropriate choice of coarse variable (order parameter), we map the underlying effective free-energy landscape for the melting of a crystalline solid. Implementation of this approach provides a means for constructing effective free-energy landscapes of structural transitions in condensed matter. The predictions of the approach for the thermodynamic melting point of a model silicon system are in excellent agreement with those of "traditional" techniques for melting-point calculations, as well as with literature values.

Accurate determination of the onset of structural transitions in complex physical systems is of crucial importance in condensed matter and materials physics. As direct access to such physical responses is typically difficult to attain experimentally, computational techniques such as molecular dynamics (MD) have provided powerful tools for probing the underlying atomic-scale dynamics and determining the transition onset. Though one of the most attractive features offered by MD lies in its ability to ultimately relate atomistic dynamics to macroscopically observable physical behavior, computing the evolution of all of the atomic coordinates over coarse (observable) time scales poses a severe limitation to the method. Recently, significant contributions have been made in addressing such time-scale limitations (see, e.g., [1-3]). Toward this goal, the so-called coarse molecular-dynamics (CMD) approach [4] was developed as an attempt to circumvent shortcomings for obtaining and analyzing the evolution of slow coarse-grained variables ("observables") of complex dynamical material systems. The projection operation formalism [5] relates, in principle, microscopic dynamics to such slow evolution; yet, exact formulas for the corresponding noise and memory terms are practically inaccessible. CMD circumvents the evaluation of such terms by estimating on the fly the thermodynamic driving forces for the slow evolution, as well as its local dynamics. This CMD approach has been used to study non-equilibrium phenomena. For example, the CMD method has been used to study the dynamics of biomolecules [4] and of water molecules filling or emptying carbon nanotubes [6]. Coarse-grained information, estimated *on-the-fly* from many short and properly initialized independent replica MD simulations, can be used to identify transition points in the physical behavior of the complex systems under consideration. CMD is a part of the so-called equation-free framework for complex/multiscale system modeling [7], which has also been used to study condensed-matter dynamical phenomena, such as line-defect motion in impure crystalline solids [8] and micelle formation [9] based on Monte Carlo (MC) simulations. This Letter aims at determining the onset of structural transitions in condensed-matter systems within the CMD framework. This is achieved by the first application of the method to predicting the melting transition of a crystalline solid through the construction of the underlying effective free-energy landscape.

The thermodynamic melting point marks the onset of the solid-to-liquid transition: this is defined as the temperature at given pressure for which the Gibbs free energies of the solid and liquid phases of a material are equal. Traditionally, computational methods for calculating thermodynamic melting points have followed two approaches: equilibrium (phase coexistence) calculations and non-equilibrium techniques. There are several phase coexistence techniques available to determine the thermodynamic melting temperature, $T_m$. Specifically, in the method developed by Landman et al. [10] solid and liquid phases in coexistence are created artificially, whereas the methods by Broughton and Li [11] and Lutsko et al. [12] are based on computing the Gibbs free energy of both solid and liquid phases as a function of temperature. On the other hand, the development of non-equilibrium methods based on MD simulations was motivated largely by the work of Galvin et al. [13] and later of others [14], which provided direct measurements of the velocity of the liquid interface of molten silicon (Si) produced during pulsed laser annealing experiments. The methods developed by Kluge and Ray [15] and by Phillpot et al. [16] are excellent examples of such non-equilibrium techniques. These techniques use a slab supercell (or a bicrystal model) containing an equilibrated solid material at a temperature below melting. The supercell is then suddenly perturbed to a temperature well above melting. This perturbation creates a melting front that is nucleated at the slab's surfaces (or the grain boundary of the bicrystal) and propagates toward the solid core at a temperature-dependent velocity. As the perturbed temperature approaches $T_m$, the melting-front propagation velocity tends to zero.

In this Letter, we use the abovementioned surface-initiated (*i.e.*, heterogeneously nucleated) melting as a representative non-equilibrium structural transition of a condensed-matter phase to demonstrate how to (i) extract the underlying effective free-energy landscape in the thermodynamic limit, and (ii) obtain the melting temperature, $T_m$, corresponding to the onset of the structural transition under consideration. This task is carried out by selecting an appropriate coarse variable, a structural order parameter that describes the state of the system, running multiple *short* MD simulations, and processing their results as described within the CMD framework.

The theoretical foundations of the method have been described in detail elsewhere [4,9]. Briefly, the method is based on the description of the evolution of the probability density, $P(\psi,t)$, where $\psi(t)$ is an appropriate coarse-grained observable that describes the state of the system. When the corresponding stochastic process is Markovian and invariant with respect to shift in time, $t$, the evolution of $P$ can be described by the Fokker-Planck equation

$$\frac{\partial P(\psi,t)}{\partial t} = -\frac{\partial}{\partial \psi}\left[v(\psi) - \frac{\partial}{\partial \psi}D(\psi)\right]P(\psi,t), \quad (1)$$

where $v(\psi) \equiv \partial <\psi(t;\psi_0)>/\partial t$ and $D(\psi) \equiv (1/2)\partial \sigma^2(t;\psi_0)/\partial t$ are the drift velocity and diffusion coefficient, respectively, and $<>$ and $\sigma^2$ denote the mean and variance, respectively. We use the equilibrium ansatz, $P_{eq}(\psi) = \exp[-G(\psi)/kT]$, where $k$ is the Boltzmann constant, $T$ is temperature, and $G(\psi)$ is the effective free energy, and integrate the equilibrium version of Eq. (1), which yields

$$\frac{G(\psi)}{kT} = -\int \frac{v(\psi')}{D(\psi')}d\psi' + \ln D(\psi) + C . \quad (2)$$

We have found the dependence of $D$ on $\psi$ to be much weaker than that of $v$ on $\psi$; consequently, in our implementation, we treat $D$ as an average value and lump the logarithmic term of Eq. (2) together with the arbitrary integration constant, $C$.

Our model consists of a slab supercell with 34 planes parallel to the surface plane, containing 50 atoms each; the free-surface planes of the slab are taken to be normal to the [001] crystallographic direction. The interatomic interactions are described by the many-body Tersoff potential (T3) for Si [17]. Although T3 overpredicts the $T_m$ of Si severely, its bond-order nature makes it an appealing functional form that is representative of a much broader class of classical force fields; within the scope of this study, however, the specific choice of interatomic potential is not significant. The equations of motion are integrated using a fifth-order Gear predictor-corrector algorithm with a time step of 0.5 fs and the temperature is kept constant by velocity rescaling at each time step. We have tested carefully this time integration scheme and concluded that it does not affect the results of this study. The order-to-disorder transition that each plane undergoes as melting proceeds is monitored by a planar order parameter, $\xi$, based on the planar structure factor [16]: $0 \leq \xi \leq 1$; $\xi = 1$ corresponds to a perfect crystalline solid plane; $\xi = 0$ corresponds to a molten plane; and $\xi = 1/2$ corresponds to the interface between the liquid and the solid. As melting proceeds, this information can be translated into the number of melted planes as a function of time, which, when normalized with the total number of planes in the system, constitutes our choice of order parameter, $\psi$, which describes properly the state of our slab system. By definition, $\psi = 0$ corresponds to a perfect crystalline solid, whereas $\psi = 1$ corresponds to a melt. For comparison purposes, we first determined $T_m$ using the method introduced by Phillpot et al. [16], where a melt/crystal propagation front is created and the temperature-dependent front propagation velocity, $v_p$, is monitored. Implementation of the method of Ref. 16 for our Si slab model involved a thicker slab (62 planes of 50 atoms each) than that for the implementation of the CMD approach, initiation at $T = 1500$ K and heating to the temperature of interest at a rate of 5 K/ps, and generation and monitoring of twenty-five propagation-front profiles at each temperature studied. This procedure gives the results of Fig. 1, where extrapolation of $v_p$ to zero yields $T_m = 2562 \pm 10$ K.

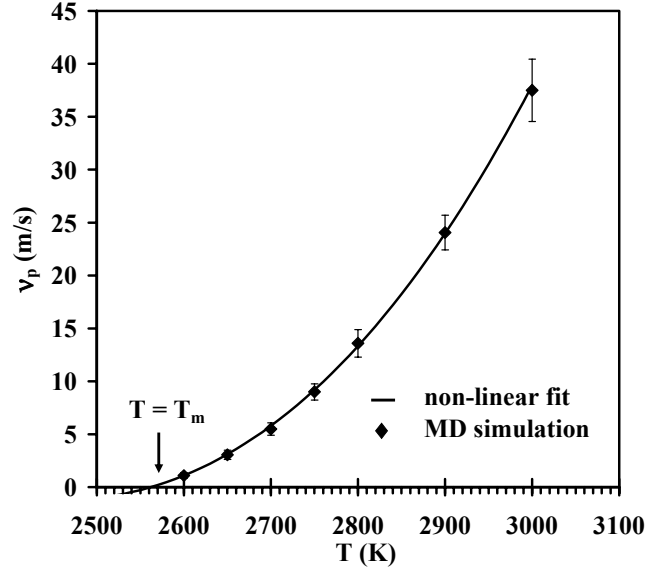

FIG. 1. Dependence of the melt/solid interface propagation velocity, $v_p$, on temperature, $T$, for a slab supercell model of crystalline Si at temperatures above melting. The symbols correspond to MD simulation results and the solid curve is a nonlinear fit to the MD results. Extrapolation to the limit $v_p \rightarrow 0$ yields a thermodynamic melting temperature, $T_m$, of $2562 \pm 10$ K. In spite of the shown trend in the error bars, the fluctuations about the mean, $<v_p(T)>$, increase substantially as $T \rightarrow T_m$.

It should be emphasized that as $T \rightarrow T_m$, the very slow interfacial propagation speed in conjunction with the increased amplitude of the fluctuations of the dynamical variables make the accurate determination of $v_p$ very demanding computationally, through analysis of extremely long MD trajectories.

We implemented the CMD approach by setting our system to the temperature of interest and using a lattice parameter corresponding to the zero-pressure isobar; it is important to keep the system at zero pressure to avoid development of thermomechanical stresses. The dependence of the lattice parameter on $T$ can be obtained by carrying out an isothermal-isobaric MD simulation [18]. At $t = 0$, the system is initialized/forced in order to satisfy a prescribed value of $\psi$. For example, choosing a value of $\psi = 4/34$ corresponds to setting an initial configuration with four melted planes. The initialization of the system satisfying this coarse description ("lifting" transformation) is non-unique [7]; in this work, it is carried out using a plane-by-plane Metropolis MC-type scheme with an added bias potential, $U_{bias}$. This has the form of a harmonic potential, $U_{bias} = K(\xi - \xi_{obj})^2$, with a stiff spring constant (we have used $K = 5 \times 10^4$) and it allows for fast sampling toward the objective values at each plane; $\xi$ denotes the running value of the planar order parameter at the plane of interest and $\xi_{obj}$ denotes its objective value. Upon successful initialization consistent with the desired $\psi$ (i.e., melting the desired number of planes and setting the new interface locations), we proceed by time stepping through MD for a 200 ps horizon, while recording points in the coarse variable trajectory, $\psi(t)$, every 0.5 ps. We repeat this process by choosing values of $\psi$ between 4/34 and 32/34 at increments of 2/34. At each initial order parameter value of choice, twenty-five independent replicas are generated and the trajectories are monitored along with their respective variance as they proceed to relax the system from its imposed initialization. By computing the slopes of the averaged coarse variable evolution and its variance, we estimate the local effective drift velocity and diffusion coefficient of Eq. (1).

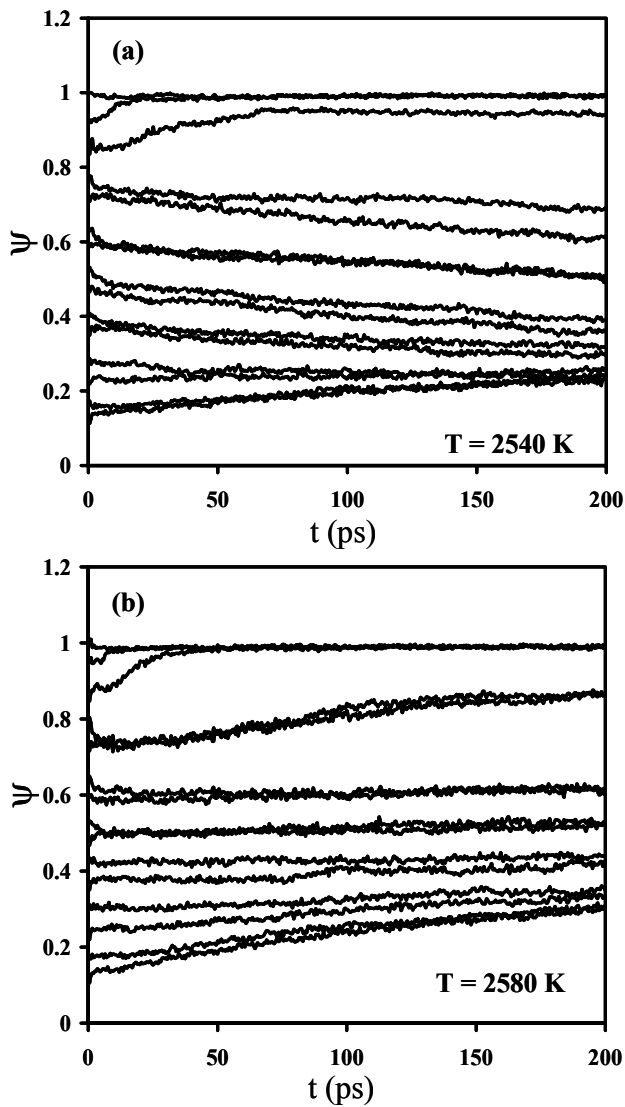

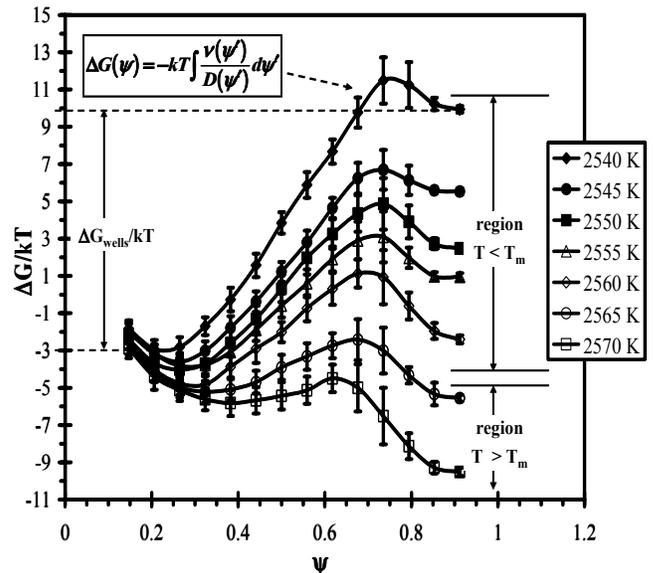

FIG. 2. Evolution of the coarse variable, $\psi(t)$, initialized at different values, $\psi_0$, for the Si slab model at temperatures of (a) 2540 K and (b) 2580 K, i.e., below and above $T_m$, respectively. The evolution exhibits drift toward the "two potential wells," the solid and the melt.

Here, we report results over the $T$ range from 2540 K to 2580 K obtained at increments of 5 K. At each $T$, we initialize the system at chosen values of $\psi$ over the interval $0 \leq \psi \leq 1$. Figure 2 shows the evolution of the coarse variable, $\psi(t)$, for various initial conditions $\psi_0 = \psi(t=0)$ at $T = 2540$ and 2580 K. In Fig. 2(a), $T = 2540$ K, it is shown that in almost all of the cases, $\psi(t)$ drifts toward a solid state characterized by $\psi < 0.5$; melting occurs only when the system is initialized at values very close to $\psi_0 = 1$. In Fig. 2(b), $T = 2580$ K, it is shown that $\psi(t)$ drifts toward a molten state, a clear indication that $T > T_m$. The slight drifting observed for $\psi = 0.6$ is due to the proximity of the system to its saddle-point configuration at this state. It is important to note that the time interval of 200 ps chosen for the MD simulations is only a small fraction of the time required for the evolution of the slowest coarse variables to attain steady state. At each $T$, analyzing the differently initialized coarse-variable trajectories yields the diffusion coefficient and the drift velocity as functions of the initial value of the coarse variable, $\psi_0$.

With $v(\psi)$ and $D(\psi)$ available, Eq. (1) can be reconstructed and integrated as given by Eq. (2) to yield the effective free-energy landscape shown in Fig. 3.

FIG. 3. Effective free-energy surface as a function of the coarse variable, $\psi$, for the Si slab model at various temperatures around the melting transition.

This landscape exhibits two (thermodynamic potential) wells: one corresponding to the solid state and another one corresponding to the molten state. The statistical uncertainty of the results of Fig. 3 is highest and lowest in the vicinity of the saddle points and minima (wells) of the effective free energy, respectively. Drawing from thermodynamic coexistence criteria, we relate the free-energy difference between the bottoms of the two wells, $\Delta G_{\text{wells}}$, (as indicated by the two horizontal dashed lines in Fig. 3) to the departure from the equilibrium phase coexistence (melting) temperature, $T_m$. Two regions are identified corresponding to temperatures above and below the thermodynamic melting temperature ($T > T_m$ and $T < T_m$, respectively), along with their corresponding activation barriers for the melting transition. A plot of $\Delta G_{\text{wells}}/kT$ as a function of $T$ is shown in Fig. 4. The temperature for which this free-energy difference goes to zero corresponds to $T_m$. A linear fit to the CMD results in conjunction with the phase coexistence criterion ($\Delta G_{\text{wells}} \to 0$) yields $T_m = 2564 \pm 2$ K.

We find the implementation of this approach to provide an accurate and computationally efficient method for constructing the effective free-energy landscapes that govern structural transitions in condensed-matter systems. More specifically, we find the predictions of this approach for the melting transition of the model silicon system used in this study to be in excellent agreement with those of "traditional" methods; these include results from analysis of *long* MD simulations, such as those shown in Fig. 1, as well as values reported in the literature for the thermodynamic melting of the Si model described by this Tersoff potential (T3) based on solid-liquid coexistence analysis: these values are 2547±22 K [19] and 2567 K [20]. More importantly, this agreement demonstrates the power of the CMD approach in establishing the connection between the non-equilibrium evolution of the coarse variable and its underlying effective free-energy gradients, as ultimately captured in the effective free-energy landscapes, such as those shown in Fig. 3. Such thermodynamic information allows for identification of important features in the landscape, which relate to the inherent stability of the system.

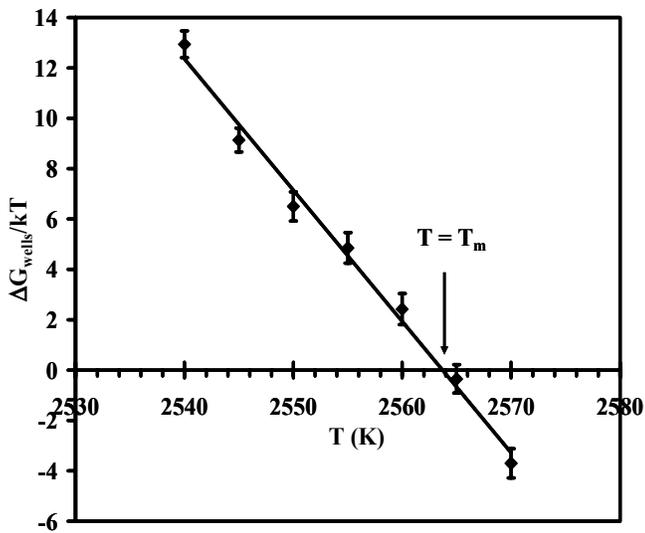

FIG. 4. Temperature dependence of the scaled free-energy difference between the two potential-well minima, yielding the melting temperature, $T_m$, for the Si slab model.

As it has been shown elsewhere [4,6,8,9], having access to the effective free energy allows for the extraction of kinetic information relevant to the rates of exchange between stable basins; however, this calculation is not shown here. It is important to note the connection between the atomistic and coarse-grained descriptions of the system, which is established through the definition of the (system-size dependent) coarse variable. Although 34 planes in the slab supercell model have been sufficient to capture the "right" transition onset for the interatomic potential employed, it is emphasized that the larger the system size (more planes), the better the coarse graining of $\psi$ and the finer the increments used when sampling its space.

In conclusion, we have demonstrated that by identifying the proper coarse variable (order parameter) that describes the state of a condensed-matter system and by performing *short* and properly initialized MD simulations within the CMD framework enables the extraction of the underlying effective free-energy landscape characteristic of heterogeneously nucleated melting of crystalline solids. In conjunction with a phase coexistence criterion, this allows for determining the onset of the melting transition expressed by the thermodynamic melting temperature. Finally, we emphasize that although the results reported in this Letter refer to the melting of a specific Si model, the CMD approach is quite general and may be helpful in determining other types of structural-transition onsets in condensed matter, including stress-induced crystalline phase transformations, as well as other order-to-disorder (*e.g.*, solid-state amorphization) and disorder-to-order (*e.g.*, crystallization) transitions; selecting appropriate coarse-grained variables is crucial to the success of this approach.

This work was supported by the National Science Foundation through Grant Nos. CTS-0205584, ECS-0317345, CTS-0417770, and by DARPA.

*Corresponding author.
Electronic address: maroudas@ecs.umass.edu